\documentclass[preprint,preprintnumbers,amsmath,amssymb]{revtex4}
\usepackage{graphicx}% Include figure files
\usepackage{dcolumn}% Align table columns on decimal point
\usepackage{bm}% bold math
\usepackage{epsfig}

\newcommand{\geqnew}{\stackrel{>}{\!\ _{\sim}}}
\newcommand{\leqnew}{\stackrel{<}{\!\ _{\sim}}}
\begin{document}

\title{Visualizing the collapse and revival of wavepackets
in the infinite square well using expectation values}

\author{R. W. Robinett} \email{rick@phys.psu.edu}
\affiliation{%
Department of Physics\\
The Pennsylvania State University\\
University Park, PA 16802 USA \\
}%

\date{\today}

\begin{abstract}

We investigate the short-, medium-, and long-term time dependence
of wave packets in the infinite square well. In addition to 
emphasizing  the  appearance of wave packet revivals,  i.e., situations 
where a spreading wave packet reforms with close to its initial shape and 
width, we also examine  in detail the approach to the collapsed phase
where the position-space probability density is almost uniformly spread
over the well.  We focus on visualizing these phenomena in both
position- and momentum-space as well as by following  the 
time-dependent expectation
values of and uncertainties in  position and momentum. We discuss
the time scales for wave packet collapse, using both an autocorrelation 
function analysis, as well as  focusing on expectation values and find two
relevant time scales which describe different aspects of the
decay phase.  In an Appendix, we briefly discuss wave packet
revival and collapse in a more general, one-dimensional power-law potential 
given by $V_{(k)}(x) = V_0|x/a|^k$ which interpolates between the 
case of the harmonic oscillator  ($k=2$) and the infinite well ($k=\infty$).

 \end{abstract}

\maketitle

\begin{flushleft}
 {\large {\bf I. Introduction}}
\end{flushleft}
\vskip 0.5cm

The study of wave packet solutions of the Schr\"odinger equation 
in model potentials can illuminate many aspects of wave mechanics,
both semi-classical features which have an obvious classical analog,
  as well as purely quantum mechanical
effects.  Since the first numerical studies of wave packet motion
\cite{goldberg_1} appeared in this journal, there have been a large
number of papers illustrating such phenomena as wave packet interactions 
with square barriers or wells \cite{barrier_1}, 
multi-well systems \cite{multi_well},
multi-dimensional scattering \cite{multi_dimensional},
and in systems of relevance to solid-state physics 
\cite{solid_state_physics}. Popular
simulation packages \cite{styer_1} now  allow students to easily
visualize the time-evolution of quantum states in a number of systems 
(as opposed to the more traditional static images of single stationary 
state solutions) and allow 
them to systematically vary the values of important physical parameters
to study the dependence on such quantities 
as the particle mass, incident energy, and barrier/well properties
such as height/depth and width.

Such studies have become increasingly relevant as pedagogical tools 
as wave packet propagation in more realistic quantum mechanical bound state 
systems has  been probed experimentally, especially the 
behavior of Coulomb wave packets
on circular \cite{circular} or elliptical \cite{elliptical}
orbits which are accessible in  Rydberg atom states
\cite{rydberg_experimental}. One of the most intriguing aspects of
such systems is that initially localized wave packets, which have
a short-term time evolution which exhibits simple classical behavior,
will spread significantly after several orbits, entering  a so-called
collapsed  phase, only to reform later in the form of
a quantum revival in which the spreading reverses itself and the
wave packet relocalizes.

One of the early papers to point out the possibility of such 
collapse/revival behavior in the context of the infinite well was
by Segre and Sullivan \cite{segre_and_sullivan}.
They studied the motion of bound 
state wave packets in such a system using a simple ``sum over energy
eigenstates''  method.  These authors noted, in passing only, 
many of the same aspects observed in the Coulomb
system, namely the spatial spreading of the wave packet, continuing
until the probability density is almost uniformly spread over the entire well,
and then reforming into something like the $t=0$ configuration. 

More recently, Bluhm, Kosteleck\'{y}, and Porter \cite{square_1}
have examined the
time evolution and revival structure of wave packets in more generic
one-dimensional systems (examining the harmonic oscillator and infinite
well in particular as special cases) while Aronstein and Stroud 
\cite{square_2} have focused on fractional revivals \cite{fractional_revivals} 
in the infinite well, also
presenting visualizations of position-space probability densities.
These studies have made extensive use of the autocorrelation 
function introduced by Nauenberg \cite{autocorrelation}, defined by
\begin{eqnarray}
C(t) & = & \int_{-\infty}^{+\infty} \psi^*(x,t)\,\psi(x,0)\,dx \nonumber \\
     & = & \int_{-\infty}^{+\infty} \phi^*(p,t)\,\phi(p,0)\,dp  
\label{autocorrelation_function} \\
     & = & \sum_{n=0}^{\infty} |a_n|^2 e^{iE_{n}t/\hbar} \nonumber 
\end{eqnarray}
where the final form is useful if the bound state wave packet is written 
in terms of energy eigenstates, $u_n(x)$,
via
\begin{equation}
\psi(x,t) = \sum_{n=0}^{\infty} a_n u_n(x) e^{-iE_n{t}/\hbar}
\end{equation}
The quantity $C(t)$, which  measures the overlap of the initial state 
wave packet (in either position- or momentum-space) with the state at 
later times,
can be used in its last form to derive information about the classical
period of motion, as well as revival and superrevival times \cite{square_1},
using information on the $n$-dependence of the $E_{n}$.  

While the autocorrelation function is a powerful tool for such formal
and general analyses \cite{square_3},
 plots of the time-dependence of $C(t)$ are
rather distant from the intuitive picture of initially 
semi-classical particle motion in the well, accompanied by wave packet
spreading, leading to a collapsed state.  Plots of position-space
probability densities, such as those in Ref.~\cite{square_2} for
isolated times, contour plots ($|\psi(x,t)|^2$ versus $(x,t)$) 
for  all $t$ up to a revival period 
(as in the 'quantum carpet' visualization in Ref.~\cite{square_5}), or
even in the form of more dynamic animations, all can give valuable information 
on the behavior of the quantum state in position-space as a function of time.

A much more direct connection, however,  to an  intuitive classical 
description involving particle trajectories can be obtained by evaluating 
the time-dependent expectation values of standard variables such as 
position and momentum,
$\langle x \rangle_t$, $\langle p \rangle_t$, which can then be compared to
familiar classical results, illustrating the expected quasi-classical 
behavior over the first few periods, but also demonstrating the 
results of purely quantum effects during the collapse and various
revival phases. Since the collapse is due to the spreading of the
wave packet, it is also natural to 'track' the behavior of the
spreads or uncertainties in the same variables, $\Delta x_t$, 
$\Delta p_t$, as functions of time, examining as well how these behave 
during the semi-classical, 
collapse, and revival phases. Taken together, these can offer a 
natural visualization and complementary description of the quantum state
over all of the natural time scales in the problem, including the 
semi-classical and collapsed phases which have  not been examined in as much
detail as the important revival structures. While the results we present here
will be evaluated numerically, our approach is the same as that taken
by Styer \cite{styer_2} in understanding wave packet propagation through
expectation values and uncertainties.

In this note, we begin by very briefly reviewing the autocorrelation
approach to the revival structure of wave packet propagation in the
infinite well in Section IIA. We illustrate this with calculations,
not only of the standard $C(t)$ over the revival period, but of a related
correlation function which provides information on the reformation of
the wave packet at half-revival periods as well as providing additional
illustrative evidence for fractional revivals. Probability
density plots, in both position- and momentum-space, are also provided to
illustrate the connections to the autocorrelation function approach
as we review and extend the results in Refs.~\cite{square_1}
and \cite{square_2},  but also for later comparison to results obtained from
expectation values.   In Section IIB we focus on the short-term evolution
of wave packets, visualizing through expectation values and uncertainties
in both $x$ and $p$, the quasi-classical propagation and the
quantum spreading leading to collapse. 
In Section IIC, we repeat this
analysis, but instead focus on the entire revival period, providing more 
graphic evidence of the reformation at the half-revival time. 
In contrast to earlier works, in Section IID, we focus on the medium-term
time dependence, emphasizing  the approach to the collapsed phase. 
We study the decay in the degree of correlation between the initial 
packet and later states (as measured by $|C(t)|$) as well as the 
approach to the quasi-uniform or 'flat' probability density phase
discussed by Segre and Sullivan \cite{segre_and_sullivan}. We find that
the time scale for the initial loss of coherence as measured by
$|C(t)|$ is simply related to the free particle spreading time, $t_0$,
while that which describes the approach to the 'flat' or truly collapsed 
state, $t_{\mbox{flat}}$ (as given by the approach of $\langle x \rangle_t$, 
$\langle p \rangle _t$, and most obviously  $\Delta x_t$ to their 
'flat' values), are, in fact,  different and we examine both in detail. 
In this regard, we find that an  approach involving expectation values 
and uncertainties provides useful information which is not encoded so 
obviously in the autocorrelation approach.

Finally, in an Appendix, we consider wave packet revivals in a more general,
one-dimensional power-law potential given by $V_{(k)}(x) = V_0 |x/a|^k$ as
this form interpolates smoothly between the case of the harmonic
oscillator ($k=2$) and the infinite well ($k = \infty$). Using simple
WKB methods for the estimation of the energy eigenvalues
(as was done in Ref.~\cite{sukhatme}) and the methods in 
Ref.~\cite{square_1}, we can explicitly exhibit the scaling of the revival 
time with $k$ which gives the familiar result for the infinite well and
which also exhibits the appropriate divergence ($T_{rev} \rightarrow \infty$
as $k\rightarrow 2$) as one approaches the harmonic oscillator case. 
For this same general system, we also evaluate the revival times for
half-wells where one introduces an infinite wall at the origin and
restricts the particle in $V_{(k)}(x)$ to have $x>0$. 
Gea-Banacloche \cite{bounce} has recently examined the case of a 
'quantum bouncing ball' in such a potential ($V(z) = mgz$ for $z>0$ 
so that $k=1$) and we find that our general result reproduces that special 
case.

\begin{flushleft}
 {\large {\bf II. Visualizing wavepacket revivals and collapse}}
\end{flushleft}
\vskip 0.5cm

While the revival structure of wave packets in the infinite well has
been discussed in detail in Refs.~\cite{square_1} and \cite{square_2}, we
will briefly recall the most important points of an autocorrelation
function analysis of this system, extending these results slightly and
providing some complementary visualizations of related topics.

\begin{flushleft}
 {\large {\bf A. Autocorrelation function analysis
}}
\end{flushleft}
\vskip 0.5cm

For definiteness, we consider the one-dimensional infinite square well 
system, with walls at $x=0,L$, for which the most general time-dependent 
solution of the Schr\"odinger  equation  (since this system does not allow 
for unbound or continuum states)
is given by the infinite discrete summation 
\begin{equation}
\psi(x,t) = \sum_{n=1}^{\infty} a_n u_n(x) e^{-iE_{n}t/\hbar}
\label{general_solution}
\end{equation}
where the (normalized) eigenstates and energy eigenvalues are, of course, 
\begin{equation}
u_{n}(x) = \sqrt{\frac{2}{L}} \sin\left(\frac{n\pi x}{L}\right)
\qquad
\qquad
\mbox{and}
\qquad 
\qquad
E_{n} = n^2\frac{\hbar^2 \pi^2}{2mL^2}
\end{equation}
Recall that the normalization of the position- and momentum-space
wavefunctions are related via
\begin{equation}
1 = \sum_{n=1}^{\infty} |a_n|^2 = \int_{0}^{L}|\psi(x,t)|^2\,dx = 
\int_{-\infty}^{+\infty} |\phi(p,t)|^2 \,dp = 1
\end{equation}
Because we are interested in evaluating expectation values of many quantities
for somewhat detailed comparison, we will be careful to ensure that any
wave packet is appropriately normalized.

Because the quantized energy eigenvalues are integral multiples of a common
value, it is easy to see that any such solution will exhibit revivals or
reformations, where $\psi(x,t+T) = \psi(x,t)$, when 
\begin{equation}
e^{-iE_nT/\hbar} = +1
\qquad
\qquad
\mbox{or}
\qquad
\qquad
T = \frac{4mL^2}{\hbar \pi}
\label{revival_time}
\end{equation}
which will be called the revival time. 
As shown in Refs.~\cite{square_2} and \cite{square_5}, the wave packet
will also reform itself at half this time as well, but in a 
possibly different location. We see this by noting that
\begin{eqnarray}
\psi(L-x,t+T/2) & = &
\sum_{n=1}^{\infty} a_n u_{n}(L-x) e^{-iE_n(t+T/2)/\hbar} \nonumber \\
& = & \sum_{n=1}^{\infty} a_n u_n(x) e^{-iE_{n}t/\hbar}\left[-\cos(n\pi)
\cos(n^2\pi)\right] \\
& = & -\psi(x,t) \nonumber 
\end{eqnarray}
This connection can be written in the form
\begin{equation}
\psi(L-x,t+T/2) = - \psi(x,t)
\qquad
\quad
\Longrightarrow
\qquad
\quad
\psi(x,t+T/2) = - \psi(L-x,t)
\label{half_revival}
\end{equation}
or
\begin{equation}
|\psi(x,t+T/2)|^2 = |\psi(L-x,t)|^2
\label{other_half_revival}
\end{equation}
so that at half the revival time later, 
 any initial wavepacket will reform itself (same shape, width, etc.),
but at a  location mirrored about the center of the well.

The momentum behavior of the system will be given by
\begin{equation}
\phi(p,t) = \frac{1}{\sqrt{2\pi \hbar}}
\int_{0}^{L} \psi(x,t) \, e^{ipx/\hbar}\, dx
\end{equation}
at all times and we can use this connection to understand the behavior
of the wave packet in momentum space at $T/2$. We note that
\begin{eqnarray}
\phi(p,t+T/2) & = & \frac{1}{\sqrt{2\pi \hbar}} 
\int_{0}^{L}\, \psi(x,t+T/2) \, e^{ipx/\hbar} \, dx  \nonumber  \\
& = &  - \frac{1}{\sqrt{2\pi \hbar}}
\int_{0}^{L}\, \psi(L-x,t) \, e^{ipx/\hbar} \, dx \nonumber \\
& = &  -e^{ipL/\hbar} \left[\frac{1}{\sqrt{2\pi \hbar}}
\int_{0}^{L} \, \psi(y,t) \, e^{-ipy/\hbar} \, dy\right] \\
& = &  -e^{ipL/\hbar} \phi(-p,t) \nonumber 
\end{eqnarray}
so that 
\begin{equation}
|\phi(p,t+T/2)|^2 = |\phi(-p,t)|^2
\label{momentum_flip}
\end{equation}
and half a revival time later the initial momentum profile is 
also reproduced,
except flipped in sign ($p \rightarrow -p$), so that the particle 
is moving in the 'other direction'. These statements are true of 
any general time-dependent solution of the form in 
Eqn.~(\ref{general_solution}) for the infinite square well.

For a typical wave packet solution, the expansion coefficients are
such that they are sharply peaked about some large 
value of $n$, say $n_0 >> 1$, so that the packet can be characterized by a 
momentum value
$p_0 \sim n_0 \pi \hbar/L$. The period of the classical motion 
will  be given by 
$\tau = 2L/v_0$ and if we associate $v_0 = p_0/m$ we find that
\begin{equation}
\tau = \frac{2L}{p_0/m} = \frac{2L^2m}{n_0\pi \hbar}
= \frac{1}{2n_0}\left[\frac{4mL^2}{\hbar \pi}\right]
= \frac{1}{2n_0} T
\end{equation}
(See also Ref.~\cite{square_1} for a more general discussion of the 
classical period.) 
We thus expect the wave packet to undergo many classical periods before
reforming itself in a revival and note that there are at least two 
important time scales in the problem, namely $T >> \tau$.

As an explicit example of a bound-state wave packet, we will use
a quasi-Gaussian form,  with expansion coefficients given by
\begin{equation}
a_n = \sqrt{\frac{\alpha \hbar \sqrt{\pi}}{L}}
e^{-\alpha^2 (p_n - p_0)^2/2} e^{-ip_nx_0/\hbar}
\label{quasi_gaussian_expansion_coefficients}
\end{equation}
where
\begin{equation}
p_n \equiv \frac{n\pi \hbar}{L}
\qquad
\qquad
\mbox{and}
\qquad
\qquad
p_0 \equiv \frac{n_0 \pi \hbar}{L}
\end{equation}
define the central momentum, corresponding to an initial speed to
the right of $v_0 \sim p_0/m$. The initial position of the wave packet is
then given by $x_0$. (This is similar to the form used in 
Ref.~\cite{segre_and_sullivan}.)

This form gives an initial wave packet which is almost Gaussian
and hence very similar in most regards to the standard analytic result
seen in many  textbooks where explicit free-particle wave packets are
constructed. (See, for example, Refs.~\cite{liboff}, \cite{robinett_gaussian}.)
For example, with this normalization, the expansion coefficients are already
almost normalized (but we explicitly ensure the proper normalization in all
of our numerical calculations) and the initial uncertainties
or spreads in the wave packet are given (to an excellent approximation) by
\begin{equation}
\Delta p = \frac{1}{\alpha \sqrt{2}}
\qquad
\qquad
\mbox{and}
\qquad
\qquad
\Delta x_0 = \frac{\alpha \hbar}{\sqrt{2}}
\end{equation}
provided  $\Delta x_0 << L$.
We recall that the corresponding free particle wave packet with Gaussian
shape given by the continuous momentum weighting
\begin{equation}
\phi(p) = \sqrt{\frac{\alpha}{\sqrt{\pi}}}
e^{-\alpha^2(p-p_0)^2/2} e^{-ipx_0/\hbar}
\end{equation}
spreads with time with a spatial uncertainty given analytically by
\begin{equation}
\Delta x_t = \Delta x_0 \sqrt{1 + \left(\frac{t}{t_0}\right)^2}
\qquad \quad 
\mbox{where}
\qquad  \quad 
t_0 \equiv m\hbar \alpha^2 = \frac{2m(\Delta x_0)^2}{\hbar}
\label{gaussian_spread}
\end{equation}
This spreading time, $t_0$,  defines a third time scale in the 
problem which we can write in the following form
\begin{equation}
\frac{t_0}{\tau} = n_0 \pi \left(\frac{\Delta x_0}{L}\right)^2
\end{equation}
for future comparison.
For our explicit calculations, we choose the numerical values
$2m = \hbar = L = 1$ to define the model system and for the specific
wave packet solution we use
\begin{equation}
\Delta x_0 = 0.05
\qquad
\longrightarrow 
\qquad
\alpha = \frac{1}{10\sqrt{2}}
\qquad
\longrightarrow
\qquad
\Delta p = 10
\label{explicit_values_1}
\end{equation}
with 
\begin{equation}
x_0 = 0.5
\qquad
\qquad
\mbox{and}
\qquad
\qquad
n_0 = 400
\label{explicit_values_2}
\end{equation}
With these values we have $T/\tau = 2n_0 = 800$ and $t_0/\tau = \pi$
so that the wave packet will spread significantly on the time scale of 
a few classical periods.

Using this explicit wave packet, we can evaluate the autocorrelation
function given by Eqn.~(\ref{autocorrelation_function}) and we plot
$|C(t)|$ versus $t$ over half a revival period in Fig.~1. (The plot is
symmetric about the half-revival time, $T/2$.) The initial decrease in
correlation for $t \geqnew 0$ is seen as the 'decay' in $|C(t)|$, while
the structures in $C(t)$ due to many of the fractional revivals  
\cite{square_2}, \cite{fractional_revivals}  are also apparent. In order
to show evidence for the reformation at $T/2$ mentioned above, we also
show in Fig.~1 a plot of a related correlation function defined by
\begin{eqnarray}
\overline{C}(t) & \equiv  & \int_{-\infty}^{+\infty}
\psi^*(L-x,t) \, \psi(x,0)\, dx \nonumber \\
& = & \left[\int_{-\infty}^{+\infty} \phi^*(-p,t)\, \phi(p,0) \,dp\right]
e^{+ipL/\hbar}
\label{anticorrelation_function}
\end{eqnarray}
which measures the correlation between the initial wave packet and later
states  which are 'flipped' about the center of the well or,
equivalently,  reappear with opposite momentum profiles. Using this form, the 
reformation at $T/2$ is clearly seen ($|\overline{C}(T/2)| =1$)
as well as additional evidence for fractional revivals.

Also using this specific wave packet as an example, we illustrate
the position-space probability densities at several times in Fig.~2.
The initial, centered wave packet at $t=0$ (solid curve) can be seen 
later at several obvious  fractional revival times. The packet at 
$t=\tau/8$ (dashed
curve) can also be seen one-half revival time later, reformed on the
other side of the well as dictated by Eqn.~(\ref{other_half_revival}). The
position-space probability density at a more 
'random' later time ($t = 124\tau$) illustrates the observation of
Ref.~\cite{segre_and_sullivan} where the position-space probability has
become almost equally distributed over the entire well; the arrow
indicates the magnitude of a truly uniform probability distribution
given by $P_{CL}(x) = 1/L$ (corresponding to the numerical values used here.)

In order to extract information
on the momentum-space wavefunctions, we can  Fourier transform
Eqn.~(\ref{general_solution}), 
 but we can also formally invert $\psi(x,t)$ to write
\begin{equation}
\phi(p,t) = \sum_{n=1}^{\infty} a_n \phi_{n}(p) e^{-iE_nt/\hbar}
\label{momentum_space_wavefunction}
\end{equation}
where the momentum-space eigenstates corresponding to the $u_n(x)$ 
are given by
\begin{equation}
\phi_n(p) = \sqrt{\frac{\hbar}{\pi L}}
\left(
\frac{p_n}{p^2 - p_n^2}
\right)
\left\{(-1)^{n} e^{ipL/\hbar} -1 \right\}
\end{equation}
so that
\begin{equation}
\langle p\rangle_t = \int_{-\infty}^{+\infty}\,p\,|\phi(p,t)|^2\,dp
\end{equation}
and similarly for other powers of momentum, which will be useful
below.

We plot, in Fig.~3, the momentum-space probability densities corresponding 
to the times in Fig.~2 and note that the wave packet is characterized by
a reversal in momentum at $T/2$ as in Eqn.~(\ref{momentum_flip}). The
initial value of the momentum spread ($\Delta p = 10$) is also recovered at
the half-revival, but the uncertainty at many later times is governed
not by the intrinsic width of the initial $\phi(p,0)$ peak, 
but rather by the 'distance' between the two sharp and well-isolated 
momentum peaks at approximately $\pm p_0$
which gives $\Delta p \approx +p_0$ for much of the collapsed phase.

While these examples are somewhat illustrative of the time-dependence of the
expectation values and uncertainties in position and momentum variables,
we now turn our attention to more detailed examinations of the behavior
of $\langle x \rangle _t$, $\Delta x_t$, $\langle p \rangle_t$, and
$\Delta p_t$ over three important time scales.

\begin{flushleft}
 {\large {\bf B. Expectation value analysis: short-term, classical
behavior}}
\end{flushleft}
\vskip 0.5cm

In many situations, the hierarchy of times given by $\tau < t_0 << T$
will hold and we will examine the time-dependence of the
wave packet propagation  over the first few classical periods. Using our
standard parameter set, we calculate the expectation values and spread
in position over a time scale equal to the first ten classical periods, 
$(0,10\tau)$ and plot the results in Fig.~4.
 The periodic and almost classical nature of the time-dependence of
$\langle x \rangle_t$ is clear (compare to Fig.~5 which gives the
purely classical result for $x(t), v(t)$ over the same period), 
but obvious differences are also present. The amplitude of the motion, 
operationally defined by half the total
``left-to-right'' distance   in $\langle x \rangle_t$ 
(which for a purely classical particle would be constant and equal to L/2) 
clearly decreases with time in the quantum case
and the reason for this is clear from the corresponding plot of the
time-dependent spatial uncertainty,  $\Delta x_t$. 
The wave packet disperses in time (away from its
initial $\Delta x_0 = 0.05$ value,  given by the horizontal dashed line)
and its width increases, initially at least, in a manner completely 
consistent with a free particle Gaussian wave packet,
as the dotted  curve which forms the upper envelope for the solid
data is given by Eqn.~(\ref{gaussian_spread}). (In this case, the importance
of including the exact quantum time-dependence and not ignoring the
dispersion \cite{ignore_dispersion} is obvious.)
The sharp 'dips' in $\Delta x_t$ correspond to 'collisions' of the 
wave packet with the walls and have been recently studied in the context of 
a single 'bounce' \cite{robinett_and_doncheski}. 
As the wave packet gets wider, the value of
$\langle x \rangle_t$ at the classical collision times decreases 
(at the right wall) or increases (at the left wall) as the
spreading  wave packet finds it more difficult to approach the walls.
Despite this obvious quantum effect,
the short-term time dependence of the system is relatively classical, 
just as in the Coulomb case.

Turning attention now to the momentum-space behavior, we can make use of
Eqn.~(\ref{momentum_space_wavefunction}) to evaluate $\langle p \rangle_t$
and $\Delta p_t$ and we plot these values in Fig.~6 over the first ten
classical periods. The correspondence between the sharp classical 
'jumps' in  the $v(t)$ versus $t$ plot in Fig.~5 and the 
$t$-dependent quantum expectation value of $p$ in Fig.~6 is clear, 
but these changes become 
increasingly 'softer' as the wave packet broadens. The plot of 
$\Delta p_t$ versus $t$ is consistent with expectations from the
'single-bounce' case considered 
in Ref.~\cite{robinett_and_doncheski}. During periods
when the packet is not involved in collisions, the momentum spread is 
initially simply given by the intrinsic width of $\phi(p)$, $\Delta p_0$,
but increases dramatically during a collision with a wall. 
At such points,  where the momentum distribution is
shifting from a single large peak at positive $p = +p_0$ to a similar one
at $p = -p_0$, the spread in momentum values is dominated by the 'distance'
between the two peaks. For example, a simple model for the momentum
distribution at the collision time might be
\begin{equation}
P_{CL}(p) = \frac{1}{2}\left[\delta(p-p_0) + \delta(p+p_0)\right]
\label{flat_momentum}
\end{equation}
which gives $\langle p \rangle = 0$ and $\Delta p = +p_0$. This behavior is
clearly seen during collisions for times satisfying $t < t_0$.
The increasing spread in the position-space wave packet 
(approaching $\Delta p \sim +p_0$) clearly implies that 
the momentum distribution comes to include significant components with both 
$p \approx \pm p_0$ for later times as well, giving the same large value of 
$\Delta p_t$ over much of the collapsed phase 
(as seen earlier in the numerical values in Fig.~3.)

\begin{flushleft}
 {\large {\bf C. Expectation value analysis: long-term revival 
behavior}}
\end{flushleft}
\vskip 0.5cm

Turning now to the long-time dependence of the wave packet, we plot, 
in Fig.~7, $\langle x \rangle _t$ versus $t$ over one entire revival
time ($T = 800 \tau$.) In the top plot of Fig.~7, 
we show $\langle x \rangle_t$ 
for the wave packet at times given by $t=(n+1/8)\tau$ and $(n+5/8)\tau$ 
as the solid and dashed curves respectively. (The $t= n\tau$
data yield the horizontal line at $\langle x \rangle_t = 0.5L$.)
The $(n+1/8)\tau$  data (solid curve) corresponds to those times at 
which the classical particle would
be one-half of its way from the center to the right-hand side and 
classically would always be located at $\langle x \rangle_t = 0.75L$, 
as indicated by the horizontal dashed line: the $(n+5/8)\tau$ case
would similarly give $\langle x\rangle_t = 0.25L$ in the classical limit.
While the wave packet data  are   initially 
consistent with these  classical results, they  quickly flatten  during 
the collapsed phase so that the
expectation value  of position at even these times is  consistent 
with $\langle x \rangle = 0.5L$. We note that at $t \approx  800\tau = T$ the
expectation values again match the classical results, which is 
our first indication of the expected revival or reformation from this 
expectation value approach.  

Similarly, we note that at
$t \approx  400\tau = T/2$, we find that the expectation values 
for position have reversed themselves, consistent with 
Eqn.~(\ref{half_revival}) for the half-revival. 
In the lower half of Fig.~7, we show results for $\langle x \rangle_t$
for {\bf all}  times in the range  $t = (116\tau,164\tau)$ 
to illustrate the fine structure apparent in the time-dependence and 
how the wave packet oscillates, with very small excursions, 
 around $\langle x \rangle _t = 0.5L$ during most of the collapsed phase.

In order to further probe the manner in which the probability distribution 
'flattens' during the collapse and reforms during a revival, 
we plot in Fig.~8 the uncertainty $\Delta x_t$
versus time over the same time periods  as in Fig.~7: we show 
values for times given by $t = n\tau$ only. To compare the results to
a putative flat probability distribution, we note that this case would
be given by
\begin{equation}
P_{\mbox{flat}}(x) = \frac{1}{L}
\end{equation}
and would yield expectation values 
\begin{equation}
\langle x \rangle_{\mbox{flat}} = \frac{L}{2}
\qquad
\quad
\langle x^2 \rangle_{\mbox{flat}} = \frac{L^2}{3}
\qquad
\quad
\Delta x_{\mbox{flat}} = \frac{L}{\sqrt{12}} = 0.288L
\end{equation}
We note that $\Delta x_t$ grows 
(consistent with Eqn.~(\ref{gaussian_spread})), but
rapidly flattens out during  much of the collapsed phase at a value 
which is indeed consistent with $\Delta x_{\mbox{flat}} = 0.288L$
over much of the entire revival period. (At the bottom of Fig~8, we
show the behavior of $\Delta x_t$ for all times in the interval
$(116\tau,164\tau)$ indicating the oscillations around the expected
'flat' result.)

The revivals at $T= 800\tau$ and 
$T/2 = 400\tau$ are also apparent here as the uncertainty returns to its
initial value, indicating a reformation of the initial wave packet, perhaps
at a different location. We also note, however, the additional feature of
quasi-revivals at $t = 200\tau$ and $600\tau$ ($T/4,3T/4$) 
where the wave packet
 returns to its initial width. To illustrate the nature of the wave packet
at these times, we have already shown in Fig.~2 the form of the packet near 
$t \approx 200\tau$. The wave packet at $t = 200\tau$ does not precisely
return to its initial state (the real part of $\psi(x,T/4)$ is larger than
$\psi(x,0)$, while the imaginary part of $\psi(x,T/4)$ vanishes), but 
does so in a way which reproduces $\Delta x_{t = T/4} = \Delta x_0$. We can
see that this is a very special case,  as the $t = \tau/8$ wave packet is
obviously split into two pieces at $t = \tau/8 + T/4$ as it is 'half way'
to being reformed at the half-revival time.  We can also note that at
$t = 100\tau, 300\tau, 500\tau$, and $700\tau$ that there are 'anti-revivals'
in that the wave packets are spread somewhat more than during the rest of
'flat phase' of the collapsed phase.  The position-space wavefunction at
$t = 100\tau$ was also  shown in Fig.~2 to illustrate this 'anti-correlation'.

We can also visualize the long-time behavior of the momentum-space variables
by plotting $\langle p \rangle_t$ and $\Delta p_t$ versus $t$ 
over one revival time
in Fig.~9 (again,  for values of $t = n\tau$.) We note that the momentum
spread quickly saturates at the 'flat' value of $\Delta p = +p_0$, except
when it returns for revivals at $t = T/2$ and $T$. 
The expectation value of
momentum reverses sign at $t = T/2$, as expected from 
Eqn.~(\ref{momentum_flip}), while returning to its original value only at the
standard revival when  $t = T$.

While we have explicitly considered only the quasi-Gaussian wavepacket
with specific numerical values here, many of the same effects are seen
quite generally. 
Changes in most of the  parameters simply
change the appropriate time scales involved. Wave packets with different
initial positions, $x_0 \neq L/2$, do not exhibit the 
special quasi-revivals at
$t = T/4,3T/4$ because they lack the required additional symmetry. Different
functional forms for the expansion coefficients (the discrete versions
of Lorentzian momentum weightings or initially flat spatial wave packets, 
for example) all exhibit similar behavior to that discussed here.

\begin{flushleft}
 {\large {\bf D. Expectation value analysis: medium-term, collapse
behavior}}
\end{flushleft}
\vskip 0.5cm

One aspect of the  behavior of wave packets in the infinite well
which has not been examined in much detail as has  the 
pattern of revivals (fractional or otherwise) is the decay phase, the
time over which the spreading of the wave packet leads to an increasing
lack of coherence (as measured by $C(t)$) and because of which the
position-space probability density  approaches the  'flat' distribution
observed by Segre and Sullivan \cite{segre_and_sullivan}. 
 One measure of a decay time scale associated with this spreading 
is given by the
scaling behavior of $|C(t)|$ at integral values of the classical period,
namely $t = n\tau$.  For the oscillator, for example, we have
$|C(t=n\tau)| = 1$ since the wave packet motion reforms exactly with the
classical oscillation frequency. For the infinite well, we find
(after a large number of numerical trials) that the  autocorrelation
function for such integral times is initially given by
\begin{equation}
\left|C(t=n\tau)\right| \sim  e^{-(n\tau/T_C)^2}
\qquad
\mbox{for}
\qquad
\qquad n\tau \leqnew T_C
\label{decay_scaling}
\end{equation}
where the collapse time, $T_C$, is given by
\begin{equation}
T_C = \frac{T}{2\pi \Delta n^2} = 4t_0 = 
4 \left(\frac{2m(\Delta x_0)^2}{\hbar}\right)
\end{equation}
and 
where $\Delta n$ is the dispersion in the $n$-distribution for the
Gaussian wave packet. 
This result may not be unexpected  since the decrease in coherence measured 
by $C(t)$ is  directly tied to the spreading of the wavepacket and 
$t_0$ is therefore  the natural time scale which arises in an autocorrelation
analysis of this type.

Examination of Fig.~8, however, suggests that there is a second scale
involved in the decay process, namely that associated with the time it
takes for the wave packet to approach the spatially flat probability
distribution. 	In order to 
better visualize the approach to the 'flat' distribution, we plot
$\Delta x_t$ versus $t$ in the range $(0,T/8=100\tau)$ for three
different values of $\Delta x_0$ in Fig.~10.  We note that in each case the 
uncertainties initially follow  the free-particle Gaussian result in 
Eqn.~(\ref{gaussian_spread}), but then level off to the same 'flat' value
of $\Delta x_{\mbox{flat}} = L/\sqrt{12}$. We can estimate the time it takes
to reach the constant value during the collapse phase by equating
\begin{equation}
\frac{L}{\sqrt{12}} = \Delta x_{\mbox{flat}} = \Delta x_{t}
= \Delta x_0 \sqrt{1+\left(\frac{t}{t_0}\right)^2}
\end{equation}
which gives
\begin{equation}
t_{\mbox{flat}} \approx  t_0 \left(\frac{L}{\sqrt{12}\Delta x_0}\right)
=
\frac{1}{\sqrt{6\pi}} \left(T t_0\right)^{1/2}
=
\frac{8}{\sqrt{12}}\left( \frac{ m L \Delta x_0}{\hbar} \right)
\label{flattening_time}
\end{equation}
which defines still another time scale, typically an intermediate scale between
$t_0$ and $T$. Numerical studies, such as in Fig.~10, indicate that the
position-space uncertainty saturates at its 'flat' value at a time given 
roughly by $2t_{\mbox{flat}}$. The three cases shown in 
Fig.~10 then illustrate that this flattening time does indeed scale with
$\Delta x_0$, in contrast to the coherence decay time scale $T_C$ which
is proportional to $\Delta x_0^2$. We can see the same flattening time
scale (and its proportionality to $\Delta x_0$) by plotting the medium-scale
time dependence of $\langle x\rangle_t$, $\langle p \rangle _t$,
and $\Delta p_t$ in Fig.~11, all of which approach the appropriate 
'flat' value with a time scale of roughly $2t_{\mbox{flat}}$. 
It is not immediately apparent how information on this important time scale
is readily obtainable from an autocorrelation analysis (for example,
as shown in Fig.~1), providing more
evidence for the usefulness of this more intuitive approach.

\begin{flushleft}
 {\large {\bf E. Conclusions }}
\end{flushleft}
\vskip 0.5cm

We have studied the time-dependence of wave packets in the infinite well, 
their collapse and revival structure, and the time scales appropriate
for each phase by examination of the expectation values and uncertainties
in the familiar variables of one-dimensional quantum mechanics, position
and momentum. The visualizations provided here complement more formal
derivations of revival structures studied using autocorrelation function
analyses. In the case of the approach to the collapsed phase of the wave 
packet approaching a 'flattened' state, this type of expectation value 
analysis very naturally yields information on the scaling properties of the 
natural time scale for the collapse, in contrast to an autocorrelation
function approach where such information is not so immediately and
intuitively obvious.

\begin{flushleft}
 {\large {\bf Acknowledgments}}
\end{flushleft}
\vskip 0.5cm
This work was supported in part by the National Science Foundation
under Grant DUE-9950702. We are grateful for helpful  comments made by
R.~Diehl, M.~Doncheski, and J.~Yeazell as well as for 
 the very useful  advice of a referee.

\newpage

\begin{flushleft}
 {\large {\bf Appendix}}
\end{flushleft}
\vskip 0.5cm

Motivated by the simple revival structure observed for the harmonic
oscillator and infinite well potentials in Ref.~\cite{square_1}, we
present here a brief analysis of the revival time for the general,
one-dimensional power-law potential
\begin{equation}
V_{(k)}(x) \equiv V_0 \left|\frac{x}{a}\right|^k
\label{general_power_law}
\end{equation}
which interpolates between these two special cases for
$k=2$ (oscillator) and $k = \infty$ (infinite square well). 
While there is really
only one independent dimensionful parameter in this potential, namely
$V_0/a^k \equiv \alpha$, we write it in this form to emphasize the
$k\rightarrow \infty$ limit where we obtain an infinite well (of width
$2a$,  compared to our earlier analysis.) A simple variation is
the 'half' power-law potential given by
\begin{equation}
      \tilde{V}_{(k)}(x)  = \left\{ \begin{array}{ll}
               V_{(k)}(x) & \mbox{for $x>0$} \\
               0 & \mbox{for $x<0$}
                                \end{array}
\right.
\label{half_well}
\end{equation}
which is then appropriate for the analysis of the 'quantum bouncing ball'
discussed recently in Ref.~\cite{bounce} when $k=1$ and
$V_0/a \equiv mg$. 

In order to make use of an autocorrelation analysis as in Ref.~\cite{square_1},
we  require the quantized energy eigenvalues,
$E_n^{(k)}$.  Since
we are dealing with situations in which $n >> 1$, it will be a good
approximation to utilize the WKB energy quantization condition to
evaluate the $E_n^{(k)}$. The standard version (which assumes
wave-function matching at linear walls via Airy functions) is given by 
\begin{equation}
\int_{-x_0}^{+x_0} 
\sqrt{
2m\left[E_n^{(k)} - V_{(k)}(x)\right]
}
\,dx
= \left(n + \frac{1}{2}\right) \hbar \pi
\end{equation}
with $n=0,1,2,...$ and where
\begin{equation}
\pm x_0 = \pm \left(\frac{E_n}{V_0}\right)^{1/k} a
\label{turning_points}
\end{equation}
are the classical turning points. This expression 
is known to be exact for the harmonic oscillator.
When the boundary conditions at a given turning point are imposed by 
infinite walls where the wavefunction must actually vanish, the appropriate
matching coefficients are $C_{L},C_{R} = 1/2$ instead of the
$C_{L},C_{R} = 1/4$ for linear walls \cite{russian_wkb},  so that for 
the infinite square well the right hand side is replaced by
$(n+1/2) \rightarrow (n+1)$ which also gives the exact answer for this 
case. (The combination of the two to describe the 'half oscillator'
where $V(x) = m\omega^2 x^2/2$ for $x>0$ and an infinite wall at the
origin is then also given exactly with these two cases.)
Using the standard WKB form for all $k< \infty$, we find the energy 
quantization condition \cite{sukhatme} 
\begin{equation}
E_n^{(k)} = \left[(n+1/2) \frac{\hbar \pi}{2a\sqrt{2m}}
V_0^{1/k} 
\frac{
\Gamma(1/k+3/2)
}{
\Gamma(1/k+1)\Gamma(3/2)
}
\right]^{2k/(k+2)}
\label{quantized_energies}
\end{equation}
(where $(n+1/2) \rightarrow (n+1)$ for $k\rightarrow \infty$.)
This reproduces the exact oscillator and infinite well examples in the
$k=2$ and $k\rightarrow \infty$ limits.  The classical periods for
a given value of $k$ and $n$ will then be given by
\begin{equation}
\tau_n^{(k)} = \frac{2\pi \hbar}{E_n^{(k)}} (n+1/2) \left(\frac{2+k}{2k}\right)
\equiv \tau(k,n)
\label{basic_period}
\end{equation}
which reduces to the familiar oscillator and infinite well results. Using 
the result of Ref.~\cite{square_1} for the revival time, namely
\begin{equation}
T = \frac{4 \pi \hbar}{d^2 E_n/dn^2} = \frac{4\pi \hbar}{E_n''}
\end{equation}
we find that the revival times are given by the simple expression
\begin{equation}
T_{rev} = T(k,n) = \left|\frac{k+2}{k-2}\right| 2n \tau(k,n)
\label{revival_period_connection}
\end{equation}
This can be confirmed by explicit evaluation of $|C(t)|$, $|\overline{C}(t)|$ 
over one revival time for any value of $k>0$ where one finds plots which 
are very similar to Fig.~1, except that the revivals are no longer exact as
they are in the case of the infinite well. This form explicitly exhibits
the divergence we expect in the oscillator case (when $k \rightarrow 2$) where
the wave packets are exactly periodic \cite{saxon}, \cite{holstein}.

We have also performed a more numerical analysis using the quasi-Gaussian
wave packet to examine the short-time decay in correlations as measured by
the decrease in the magnitude of $|C(t)|$. When we evaluate the 
autocorrelation function at integral values of the classical period, 
we again find that
\begin{equation}
|C(t=n\tau)| \sim  e^{-(n\tau/T_C)^2}
\end{equation}
where
\begin{equation}
T_C \equiv 
\left[
\left|
\frac{k+2}{k-2}
\right|
2n \tau(k,n)
\right]
\frac{1}{2\pi \Delta n^2}
= \frac{T(k,n)}{2\pi \Delta n^2}
\end{equation}
which reproduces the result for the infinite well we have considered
more explicitly,  as well as exhibiting  the appropriate divergence
as $k \rightarrow 2$ as we approach the exactly periodic oscillator
case. This result also holds for other, non-Gaussian wave packet expansions,
with the numerical factor $2\pi$ replaced by different numerical coefficients
of the same order. Two open questions which would be interesting to explore
further in this context of the general well are whether during the 
collapsed phase the
position-space probability naturally approaches a semi-classical
probability density 
(as in Ref.~\cite{robinett_classical_probability_density}) over much of the
collapsed phase and, if so, how the 'flattening' time to approach this
semi-classical distribution scales with $k$ and other parameters, i.e., what
is the generalization of Eqn.~(\ref{flattening_time}).

The case of the 'half' general-power law potential given by 
Eqn.~(\ref{half_well}) is easily analyzed in the same manner. The WKB
energy quantization condition is applied at the classical turning points
$x=+x_0$ and $x=0$ with the result
\begin{equation}
\tilde{E}_n^{(k)} = \left[(n+3/4) \frac{\hbar \pi}{a\sqrt{2m}}
V_0^{1/k} 
\frac{
\Gamma(1/k+3/2)
}{
\Gamma(1/k+1)\Gamma(3/2)
}
\right]^{2k/(k+2)}
\label{half_quantized_energies}
\end{equation}
with the factor $(n+3/4) \rightarrow (n+1)$ for $k\rightarrow \infty$.
(Recall the discussion after Eqn.~(\ref{turning_points}) about the
appropriate matching coefficients for infinite wall boundaries.)
The classical period is then given by Eqn.~(\ref{basic_period})
with $E_n^{(k)} \rightarrow \tilde{E}_n^{(k)}$ and 
$(n+1/2) \rightarrow (n+3/4)$. We find that the same relationship 
between revival time and classical period as in 
Eqn.~(\ref{revival_period_connection}) holds and we note that our 
general  result then reproduces the revival time found in 
the special case of the 'quantum bouncer' considered 
in Ref.~\cite{bounce} when $k =1$ and $V_0/a = mg$.

Gea-Banacloche \cite{bounce} has also 
discussed a collapse time for the case of the
'quantum bouncing ball' (the $k=1$ case of the `half-well') and we note
in passing that his expression is close in spirit to the 'flattening' time
discussed for the infinite well as his expression for the collapse time scales
as $1/\Delta n$ as does $t_{\mbox{flat}}$, and not, for example,
like $T_C \propto 1/\Delta n^2$ for the decay time obtained by the scaling
properties of the autocorrelation function.

\newpage
\begin{flushleft}
{\Large {\bf 
Figure Captions}}
\end{flushleft}
\vskip 0.5cm
 
\begin{itemize}
\item[Fig.\thinspace 1.] Plot of the autocorrelation function,
$|C(t)|$,  (in Eqn.~(\ref{autocorrelation_function})) for the infinite
square well. The modulus of $C(t)$ is evaluated at integral values
of $t/\tau$ over the first half revival time; the curves are symmetric
about $t=T/2 = 400\tau$. The vertical dashed lines indicate
values of some of the possible fractional revivals, namely $(p/q)T$ for
integral values of $p<q$. The standard wave packet parameters
in Eqns.~(\ref{quasi_gaussian_expansion_coefficients}), 
(\ref{explicit_values_1}), and (\ref{explicit_values_2}) are used. The related
correlation function, $\overline{C}(t)$, 
 defined in Eqn.~(\ref{anticorrelation_function}),  is also plotted
and shows much of the same structure, including the `anti-revival' at
$t=T/2=400\tau$ where the wave packet reforms, but with  opposite momentum
values.
\item[Fig.\thinspace 2.] Position-space probability distributions,
$|\psi(x,t)|^2$ versus $x$,  for the quasi-Gaussian wave packet at various
times. The top panel shows the wave packet at and just after the initial
time. The $t=100\tau$ plot is typical of a fractional revival where the
wave packet reforms into a small number of recognizable features. The
$t=124\tau$ is typical of a more random later time when the wavefunction
approaches the 'flat' probability distribution,  $P_{\mbox{flat}}(x) = 1/L$,
indicated by the horizontal arrow. 
The $t\approx 200 \tau$ cases show a special fractional revival where an
initially central ($x_0=0$) wave packet reforms with the initial  value of
$\Delta x_0$. Finally, the reformation 
of the $t=0$ wave packet at the half-revival time, $t=T/2=400\tau$ is 
indicated, as is the reformation of the $t=\tau/8$ packet, but flipped
about the center of the well as in Eqn.~(\ref{other_half_revival}).
\item[Fig.\thinspace 3.] Momentum-space probability distributions,
$|\phi(p,t)|^2$ versus $p$, corresponding to the sample times shown in
Fig.~2; the vertical dashed lines indicate the values
$p = \pm p_0 = \pm n_0 \pi \hbar/L = \pm 400\pi$. Note the reversal
in momentum values at the half-revival time, $t = T/2 = 400\tau$. 
\item[Fig.\thinspace 4.] Average value for position, 
$\langle x \rangle_t$ (top), and spread in position, $\Delta x_t$ (bottom), 
for the quasi-Gaussian wavepacket defined by the parameters in 
Eqns.~(\ref{quasi_gaussian_expansion_coefficients}),
(\ref{explicit_values_1}), and (\ref{explicit_values_2})
 over a time interval corresponding to the
first ten classical periods. In the bottom figure, the first three
multiples of the spreading time, $t_0$, are shown (vertical dashed lines).
The horizontal dashed line corresponds to the initial spread, 
$\Delta x_0 = 0.05$. The dotted curve is given by the expression for the
time-dependent $\Delta x_t$ for free-particle Gaussian wavepackets given
by Eqn.~(\ref{gaussian_spread}).
\item[Fig.\thinspace 5.] Classical motion ($x(t)$ and $v(t)$ versus $t$) 
of a point particle in an infinite square well potential, 
starting at the center of the well  with constant speed $v_0$, moving  in the 
positive-$x$ direction.
\item[Fig.\thinspace 6.] Same as Fig.~4, but for the average value of
momentum, $\langle p \rangle_t$ (top), and spread in momentum,
$\Delta p_t$ (bottom). In the bottom plot, the initial momentum spread, 
$\Delta p_0$, is shown 
as a horizontal dashed line,  as is the maximum value, $\Delta p = +p_0$,
corresponding roughly to two equal peaks in $|\phi(p,t)|^2$
at $ p = \pm p_0$ as in Eqn.~(\ref{flat_momentum}).
\item[Fig.\thinspace 7.] Average value of position, $\langle x \rangle _t$
versus $t$, over one revival time ($T = 800\tau$). The average value of
position for $t=(n+1/8)\tau$ (solid)  and
$(n+5/8)\tau$ (dashed) are  plotted in the top figure. (For values
given by $t=n\tau$, one obtains the horizontal solid line corresponding
to $\langle x \rangle_t = 0.5L$.) The revival at
$t = T = 800\tau$ where expectation values return to their $t \approx 0$
values is evident, as is the half-revival at $t = T/2 = 400\tau$ where
the expectation values are mirrored about the center of the well as
predicted by Eqn.~(\ref{half_revival}). The value of $\langle x \rangle _t$
for all times during the interval $(116\tau,164\tau)$ is  shown
on the bottom plot, 
indicating the small oscillations about the 'flat' value of
$\langle x \rangle = 0.5L$ during most of the collapsed phase.
\item[Fig.\thinspace 8.] Uncertainty or spread in position, $\Delta x_t$
versus $t$ (evaluated at times given by $t = n\tau$) over one
revival time ($T = 800\tau$). The 'flat' value of $\Delta x_{\mbox{flat}}
= L/\sqrt{12} \approx 0.288L$ is shown as the dashed horizontal line.
The spread in position initially follows the dashed curve (indicated by
$\Delta x_{\mbox{free}}$) given by the free-particle expression in 
Eqn.~(\ref{gaussian_spread}), but then 'turns over' and approaches the
'flat'  value. 
The time interval $(116\tau,164\tau)$ is also shown to indicate the
oscillations around the 'flat' value during the collapsed phase. 
The full revival at $t = T
= 800\tau$ as well as the half-revival at $t = T/2= 400\tau$ are evident,
as are the special quasi-revivals at $T/4 = 200\tau$ and $3T/4 = 600\tau$
which appear in the $x_0 = L/2$ case only.
\item[Fig.\thinspace 9.] Average value ($\langle p \rangle_t$ versus $t$)
and uncertainty ($\Delta p_t$ versus $t$) in momentum for the quasi-Gaussian
wave packet over one revival time. The reversal in momentum seen in 
$\langle p \rangle_t$ and the reformation of the wave packet to its 
initial width  are obvious at the half-revival  time, $t =T/2 = 400\tau$.
The spread in momentum corresponding to the
'flat' distribution of probability in position-space ($\Delta p = +p_0$)
is also evident over a  large fraction of the collapsed phase.
\item[Fig.\thinspace 10.] Plot of $\Delta x_t$ versus $t$ showing the
approach to the 'flat' distribution for various values of the initial
spread, $\Delta x_0$. In each case, the curves  initially follow
the free-particle Gaussian expression in Eqn.~(\ref{gaussian_spread}),
but then 'turn over' and saturate at $\Delta x = L/\sqrt{12}$.
\item[Fig.\thinspace 11.] Plots of $\langle x \rangle_t$ (top),
$\langle p \rangle_t$ (middle), and $\Delta p_t$ (bottom) showing the
approach to the values of these quantities corresponding to the
'flat' distribution. Note that the characteristic time  for
flattening is the same in all three cases  and that it scales as indicated
in Eqn.~(\ref{flattening_time}), namely $t_{\mbox{flat}} \propto \Delta x_0$.
\end{itemize}

\newpage

\noindent
\hfill
\begin{figure}[hbt]
\, \hfill \,
\begin{minipage}{0.7\linewidth}
\epsfig{file=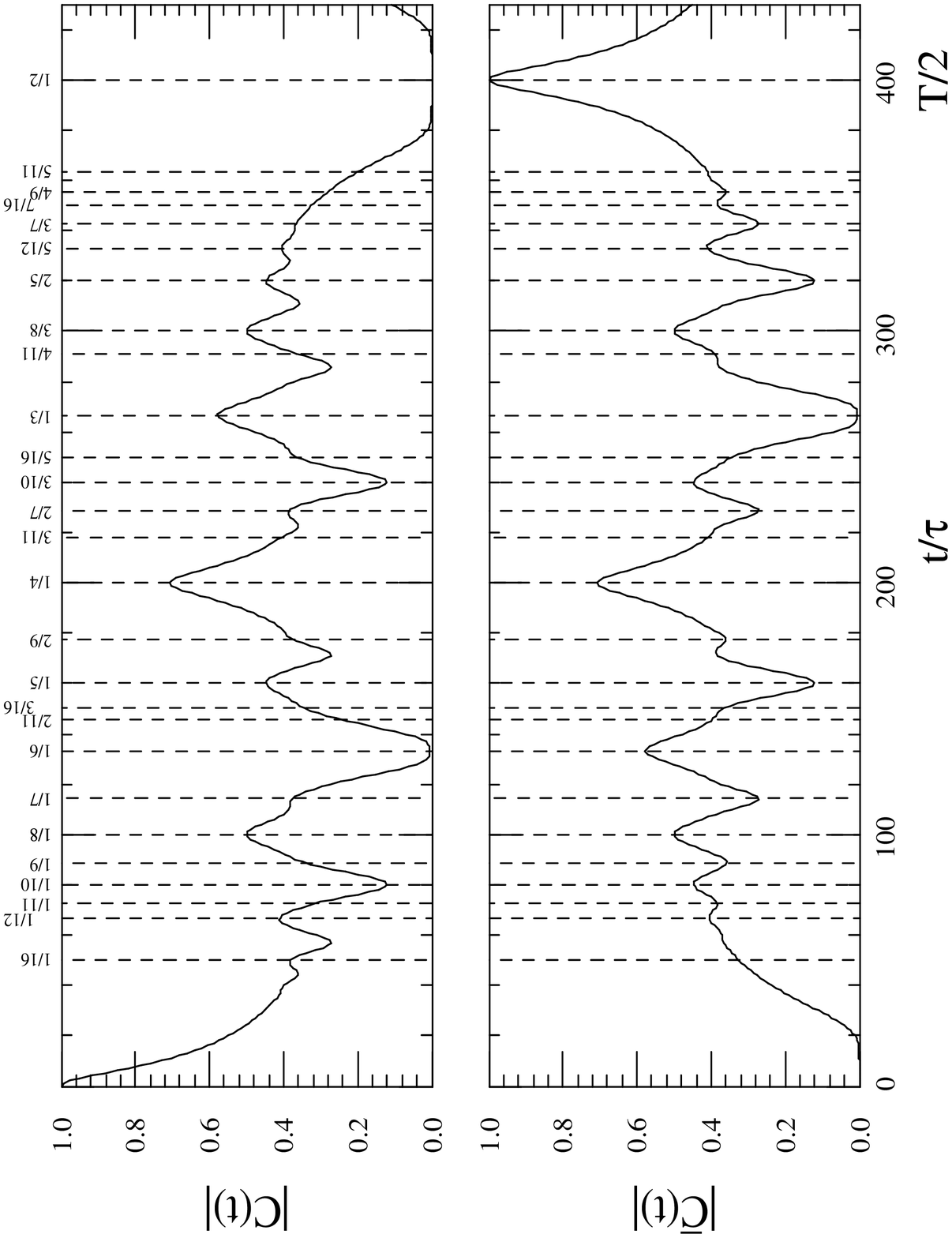,width=\linewidth}
\caption{}
\end{minipage}
\, \hfill \,
\end{figure}
\hfill 

\newpage

\noindent
\hfill
\begin{figure}[hbt]
\, \hfill \,
\begin{minipage}{0.7\linewidth}
\epsfig{file=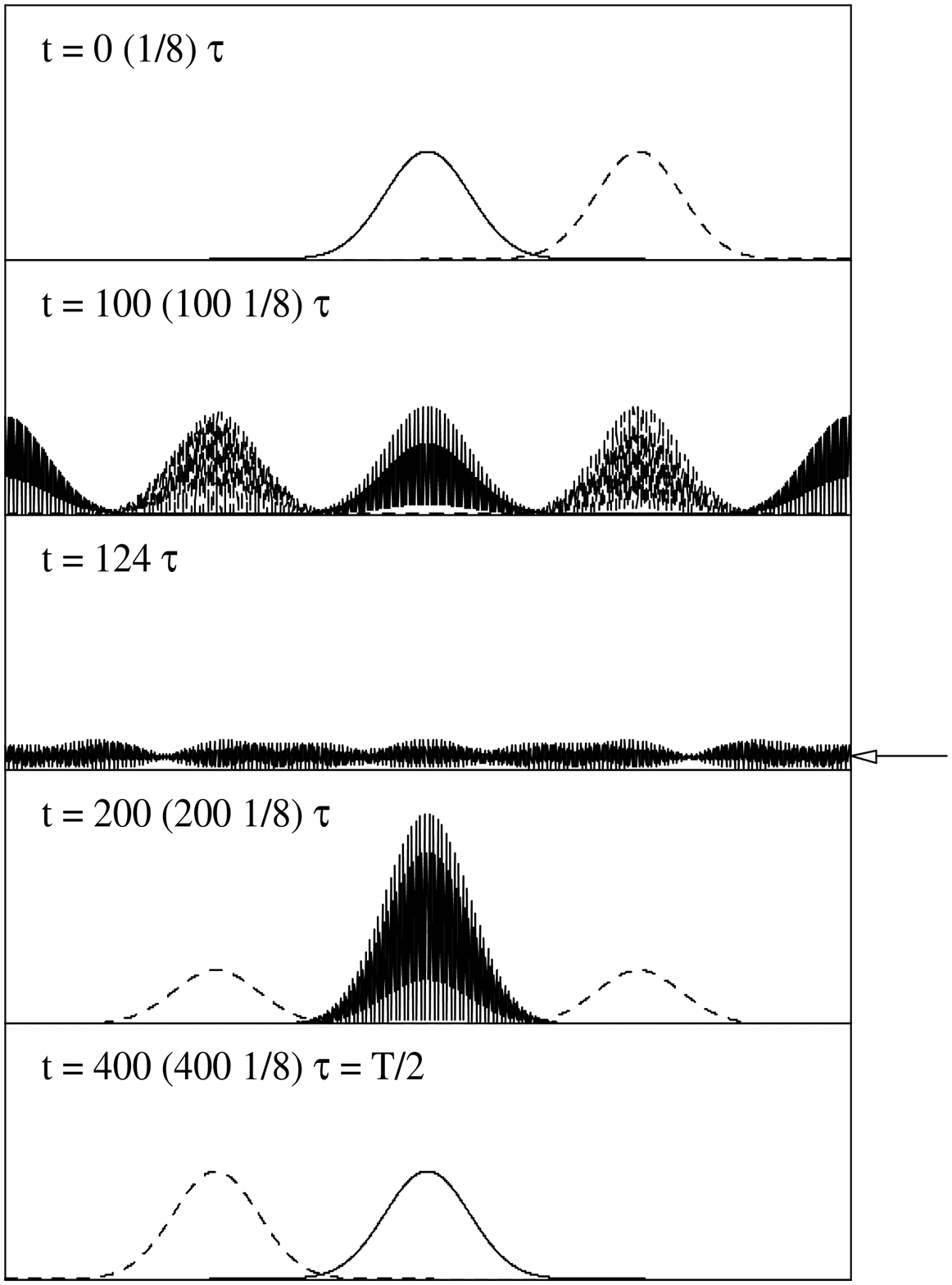,width=\linewidth}
\caption{}
\end{minipage}
\, \hfill \,
\end{figure}
\hfill 

\newpage

\noindent
\hfill
\begin{figure}[hbt]
\, \hfill \,
\begin{minipage}{0.7\linewidth}
\epsfig{file=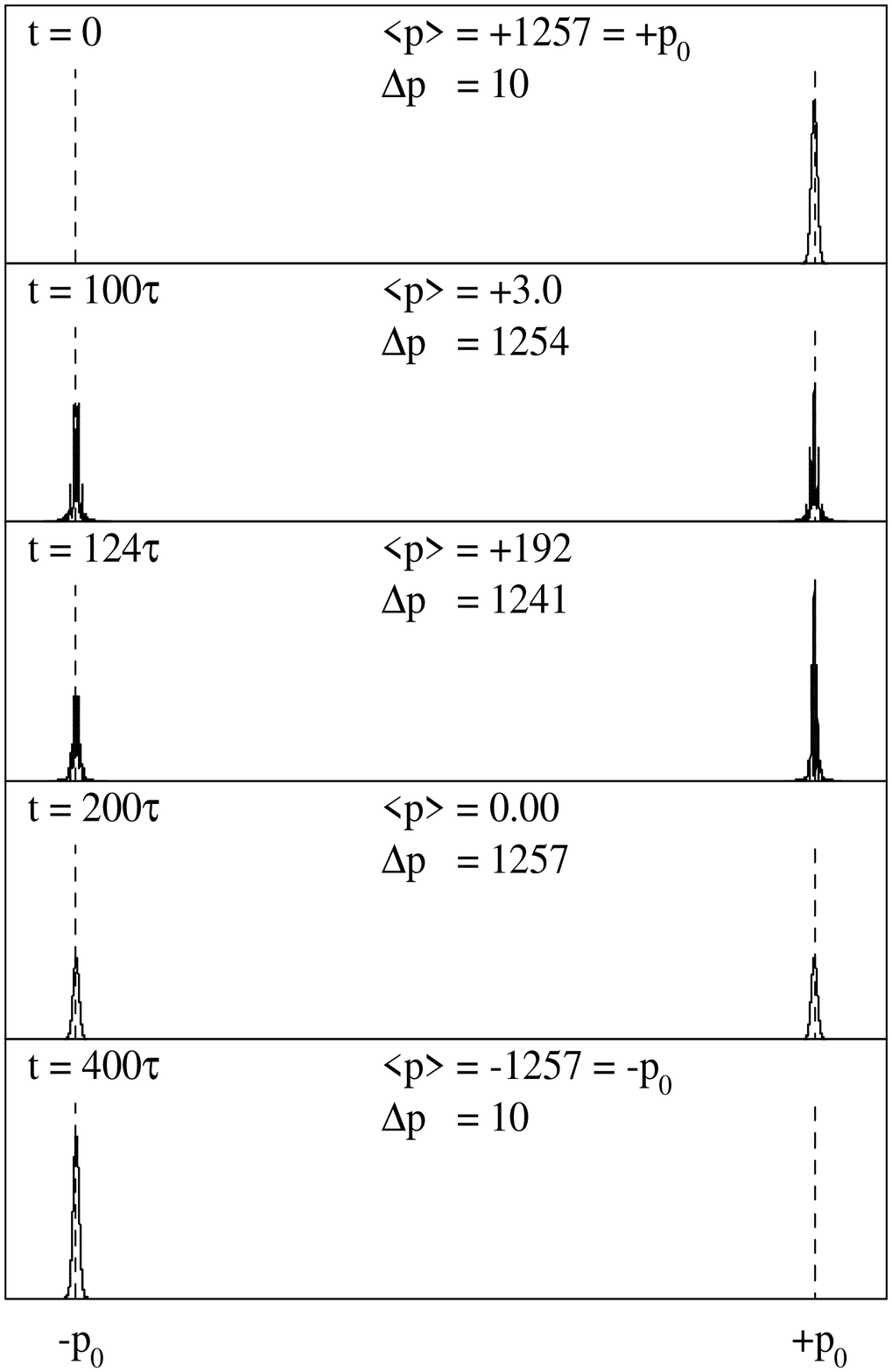,width=\linewidth}
\caption{}
\end{minipage}
\, \hfill \,
\end{figure}
\hfill 

\newpage

\noindent
\hfill
\begin{figure}[hbt]
\, \hfill \,
\begin{minipage}{0.7\linewidth}
\epsfig{file=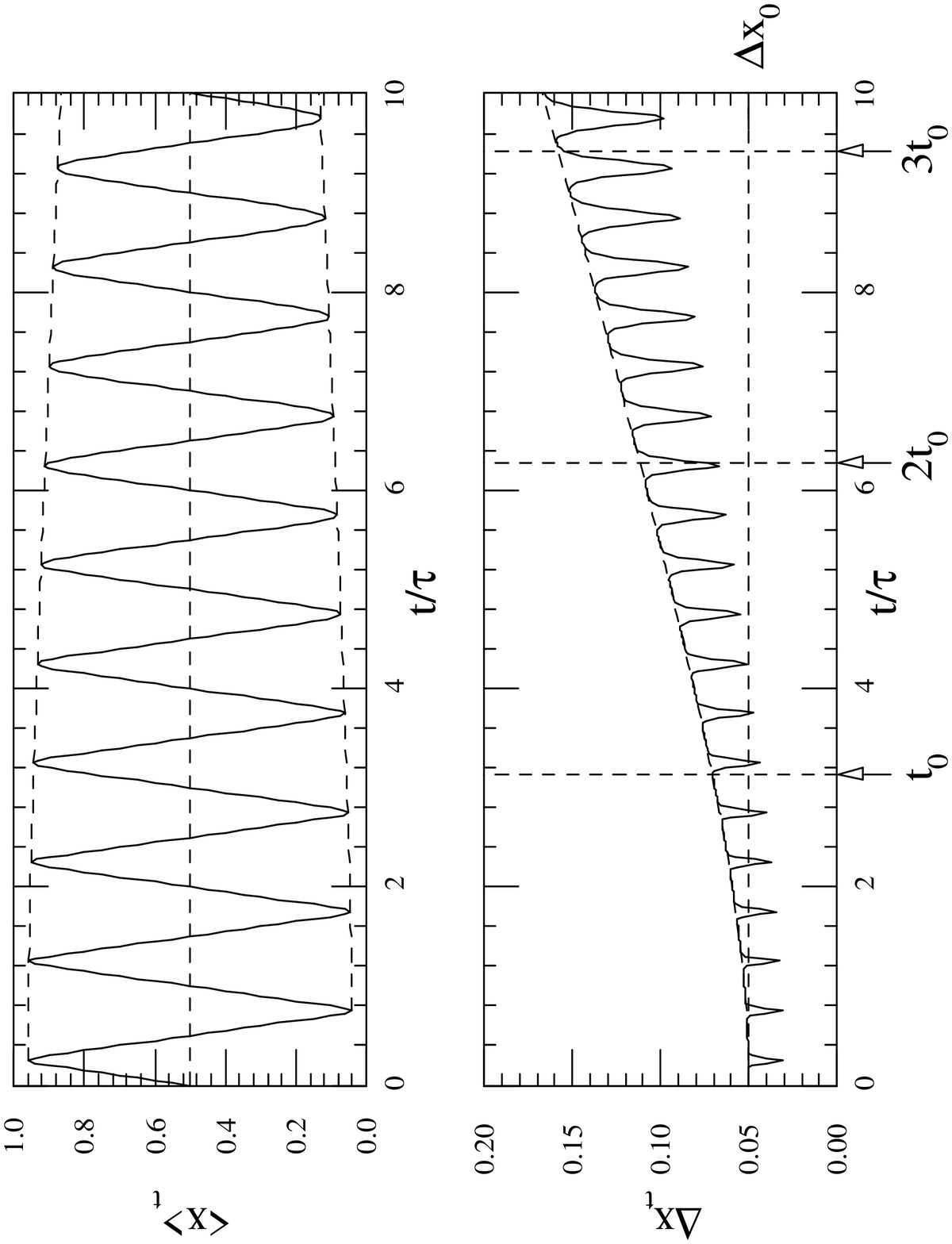,width=\linewidth}
\caption{}
\end{minipage}
\, \hfill \,
\end{figure}
\hfill 

\newpage

\noindent
\hfill
\begin{figure}[hbt]
\, \hfill \,
\begin{minipage}{0.7\linewidth}
\epsfig{file=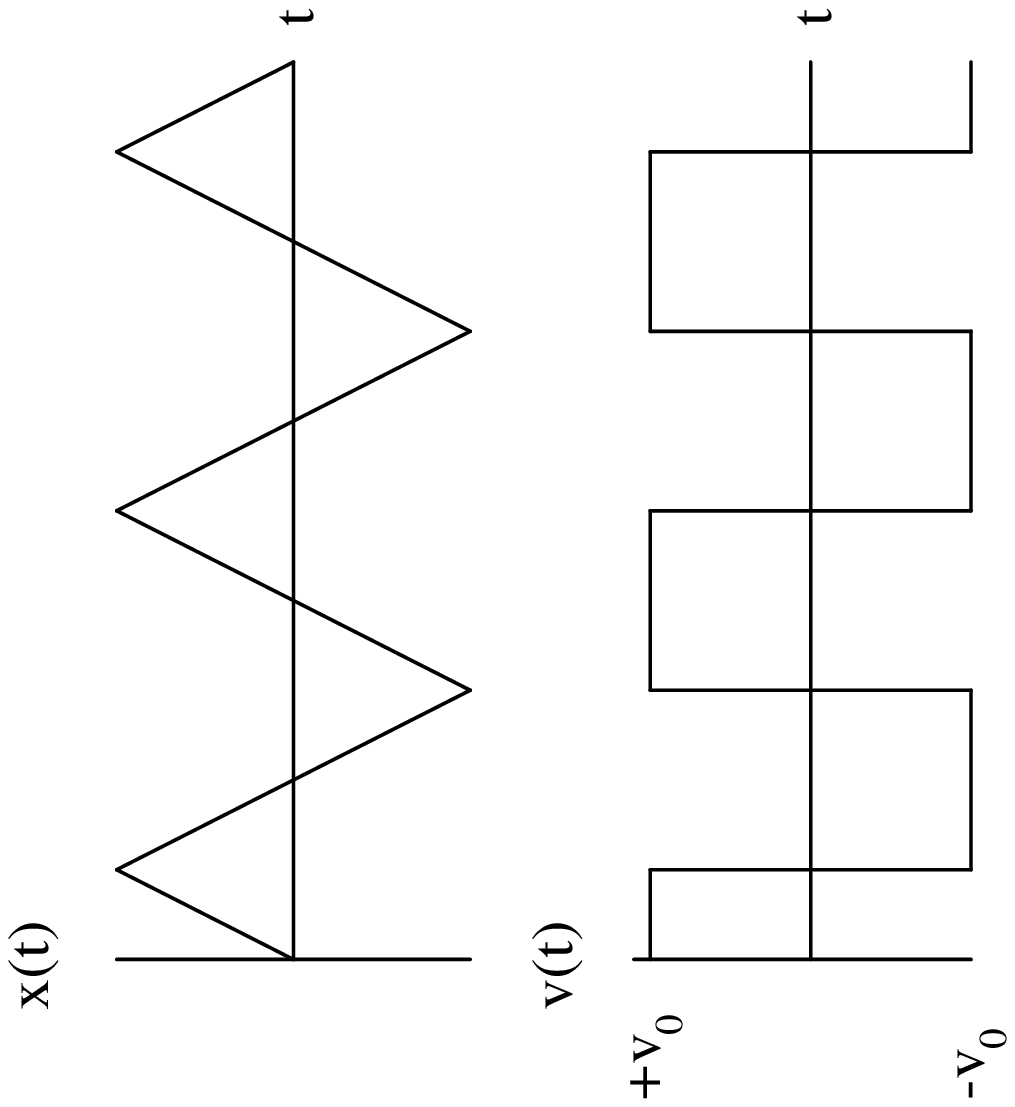,width=\linewidth}
\caption{}
\end{minipage}
\, \hfill \,
\end{figure}
\hfill 

\newpage

\noindent
\hfill
\begin{figure}[hbt]
\, \hfill \,
\begin{minipage}{0.7\linewidth}
\epsfig{file=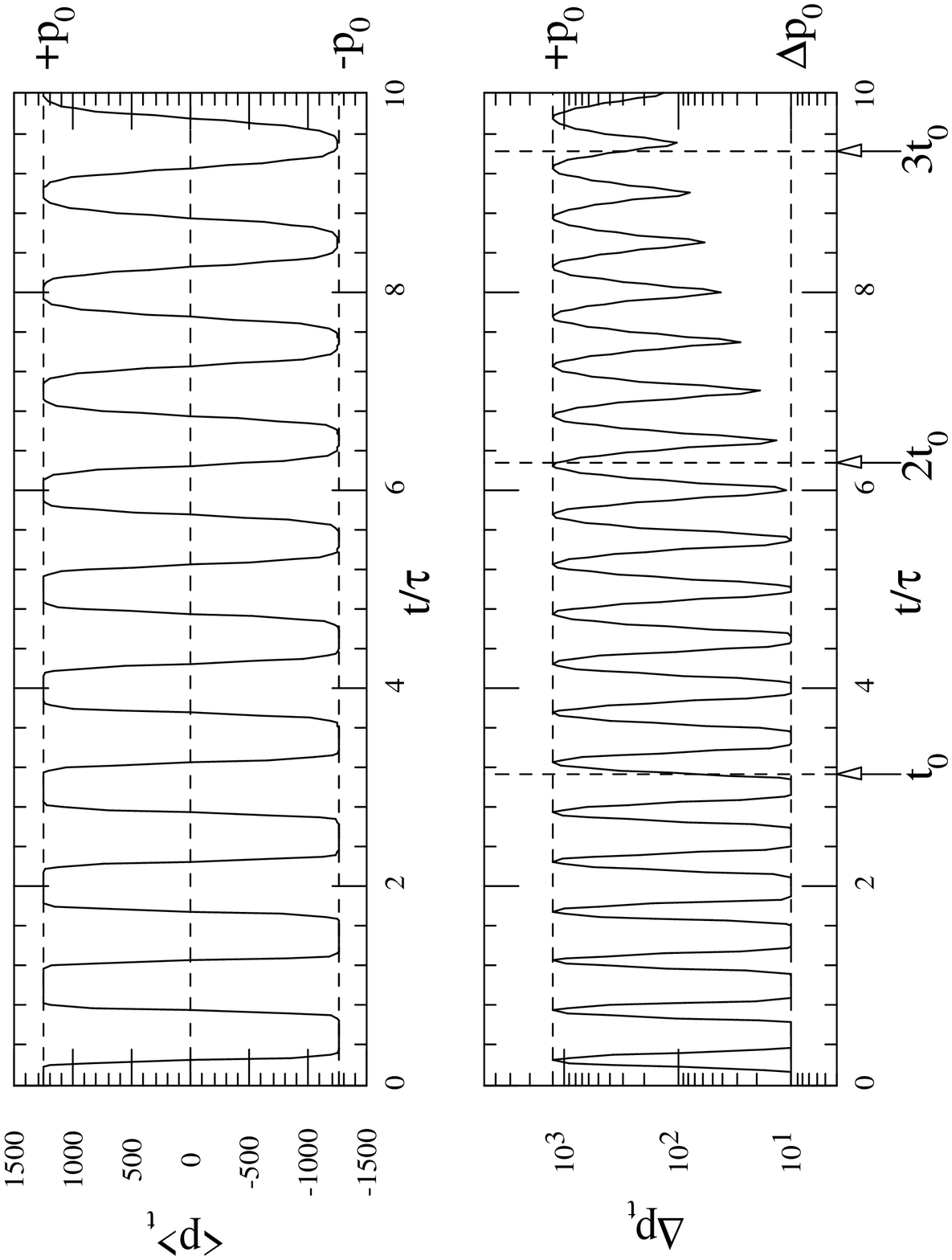,width=\linewidth}
\caption{}
\end{minipage}
\, \hfill \,
\end{figure}
\hfill 

\newpage

\noindent
\hfill
\begin{figure}[hbt]
\, \hfill \,
\begin{minipage}{0.7\linewidth}
\epsfig{file=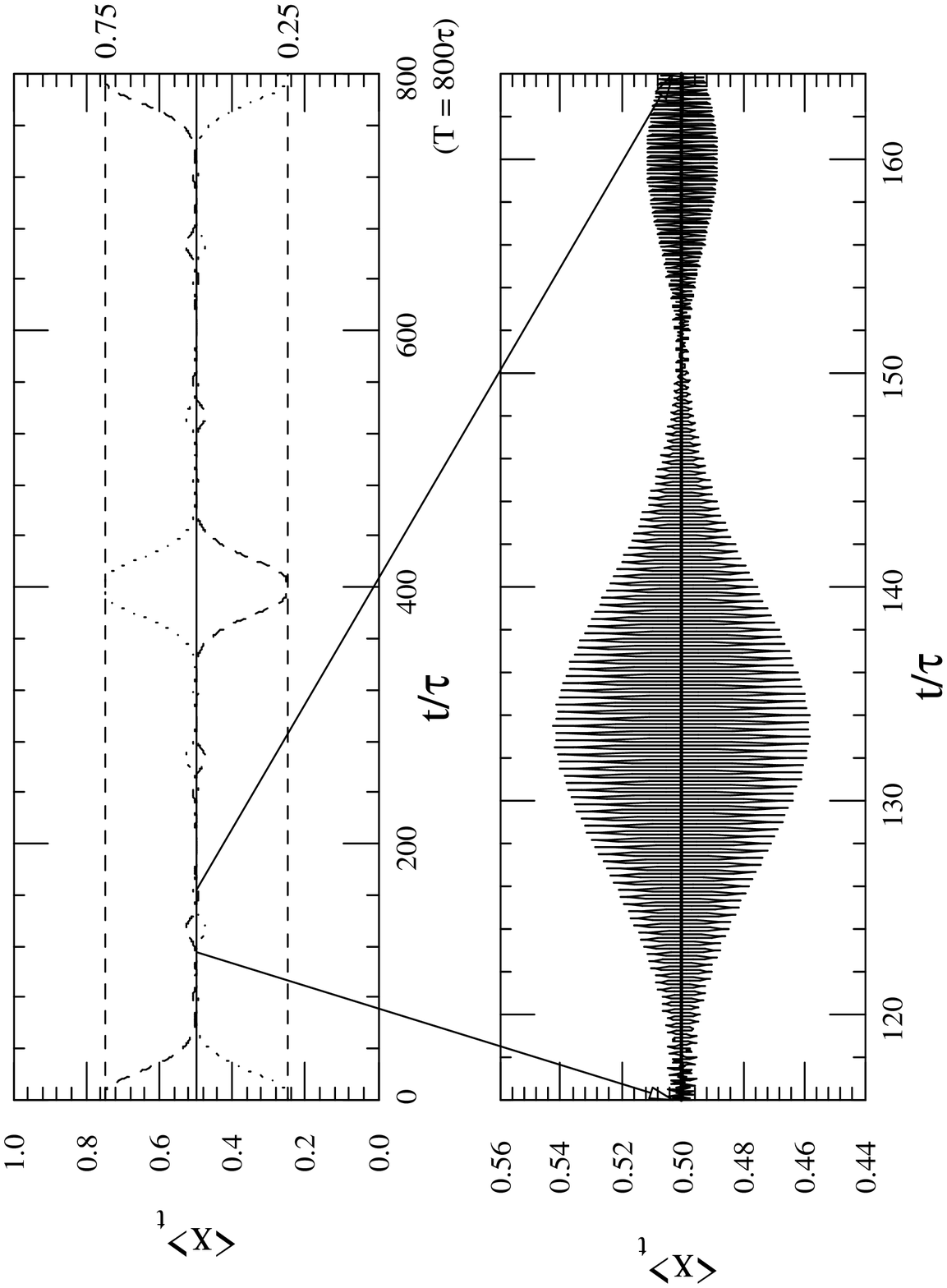,width=\linewidth}
\caption{}
\end{minipage}
\, \hfill \,
\end{figure}
\hfill 

\newpage

\noindent
\hfill
\begin{figure}[hbt]
\, \hfill \,
\begin{minipage}{0.7\linewidth}
\epsfig{file=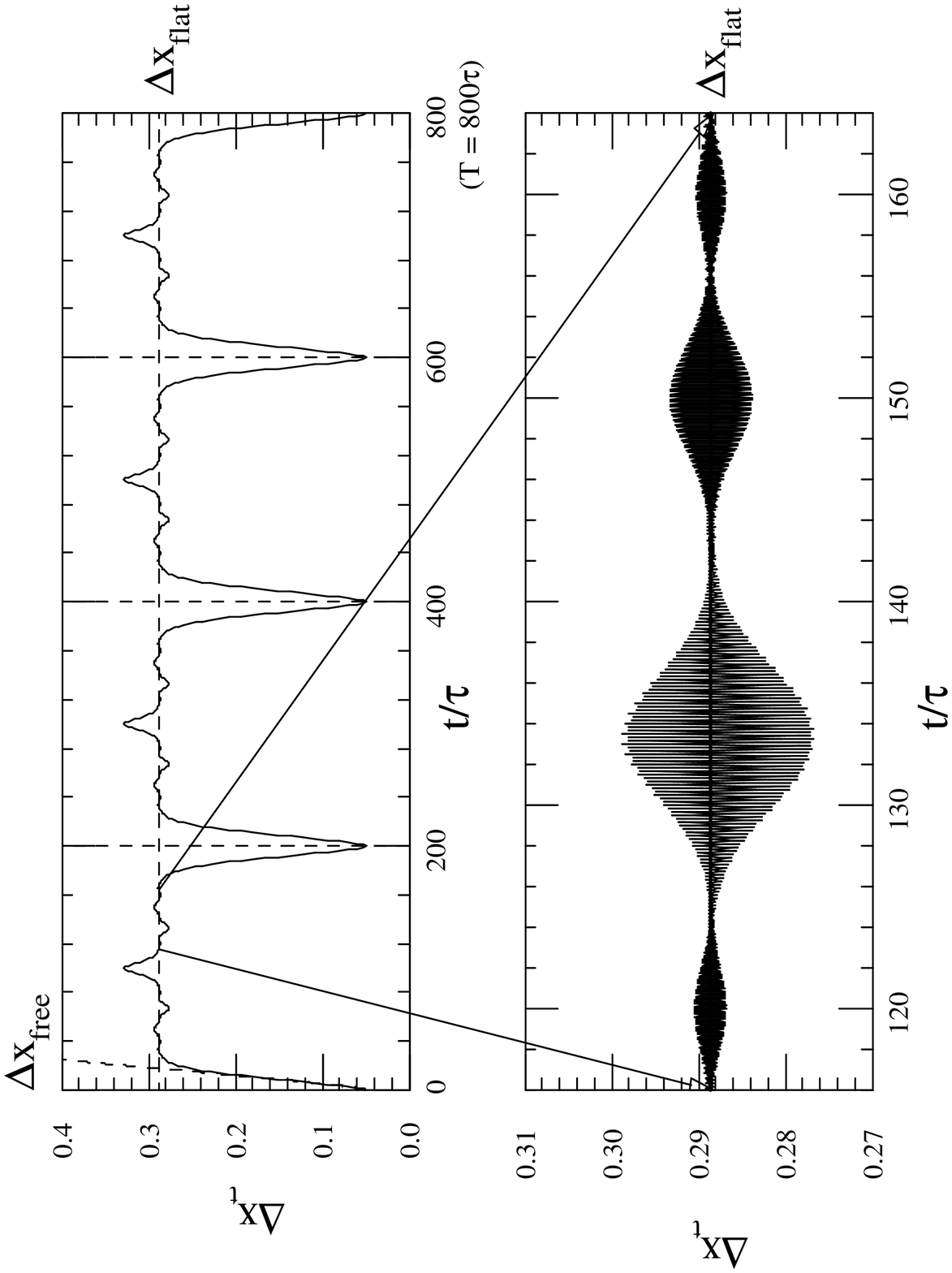,width=\linewidth}
\caption{}
\end{minipage}
\, \hfill \,
\end{figure}
\hfill 

\newpage

\noindent
\hfill
\begin{figure}[hbt]
\, \hfill \,
\begin{minipage}{0.7\linewidth}
\epsfig{file=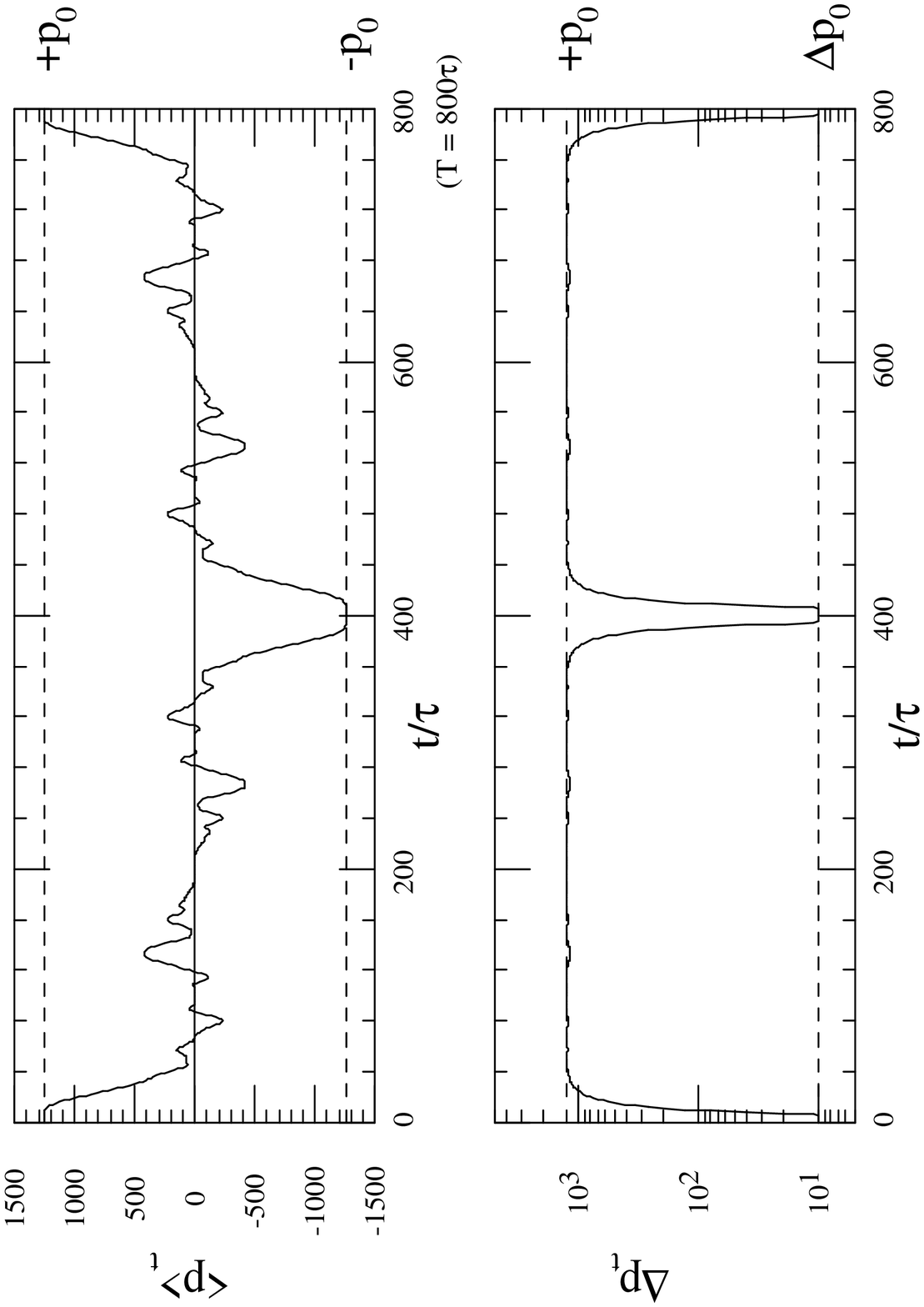,width=\linewidth}
\caption{}
\end{minipage}
\, \hfill \,
\end{figure}
\hfill 

\newpage

\noindent
\hfill
\begin{figure}[hbt]
\, \hfill \,
\begin{minipage}{0.7\linewidth}
\epsfig{file=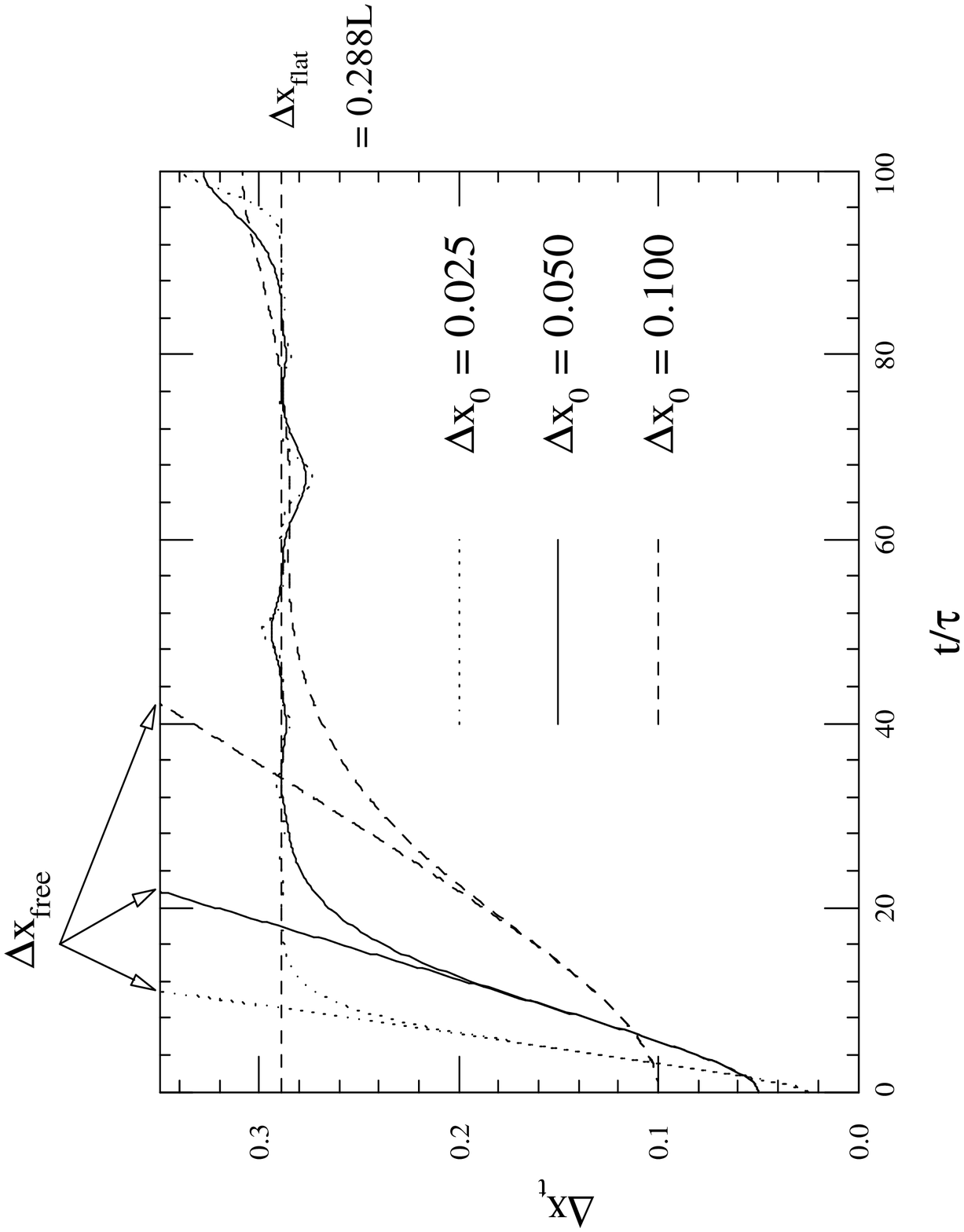,width=\linewidth}
\caption{}
\end{minipage}
\, \hfill \,
\end{figure}
\hfill 

\newpage

\noindent
\hfill
\begin{figure}[hbt]
\, \hfill \,
\begin{minipage}{0.7\linewidth}
\epsfig{file=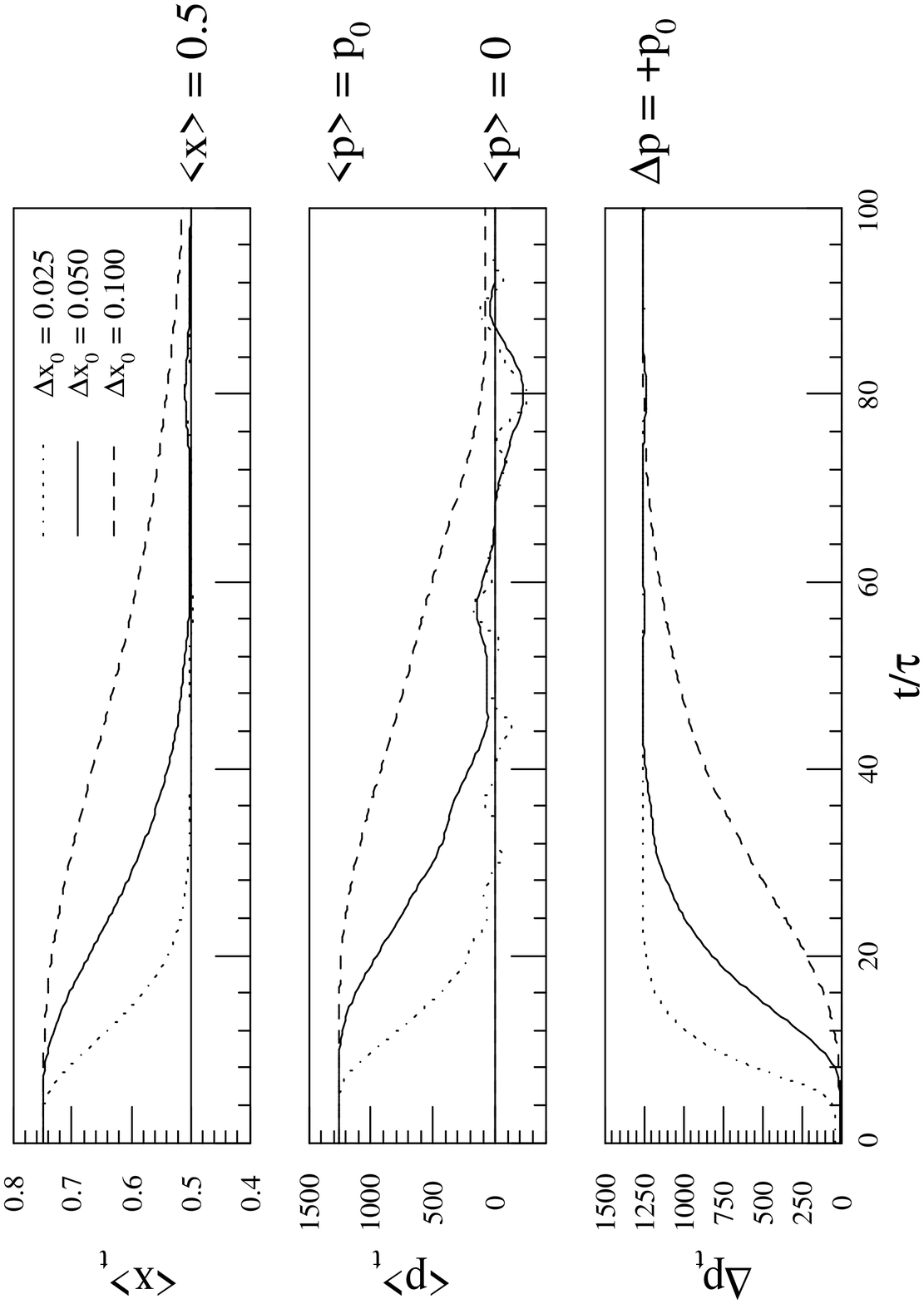,width=\linewidth}
\caption{}
\end{minipage}
\, \hfill \,
\end{figure}
\hfill

\end{document}